\documentclass[conference]{IEEEtran}

\usepackage[T1]{fontenc}
\usepackage[utf8]{inputenc}
\usepackage{newtxtext}
\usepackage[amsthm]{newtxmath}

\usepackage{amsmath,mathtools,bm}
\usepackage{dsfont}
\usepackage{relsize}

\usepackage{graphicx}
\usepackage{booktabs}
\usepackage{adjustbox}
\usepackage{float}
\usepackage{tikz}
\usetikzlibrary{arrows.meta,positioning}

\usepackage[dvipsnames,table]{xcolor}
\usepackage[numbers,sort&compress]{natbib}

\usepackage[hidelinks]{hyperref}

\usepackage{tfrupee}
\usepackage{enumitem}
\usepackage{setspace}

\theoremstyle{plain}
\newtheorem{theorem}{Theorem}
\newtheorem{lemma}{Lemma}
\newtheorem{proposition}{Proposition}
\newtheorem{corollary}{Corollary}

\theoremstyle{definition}
\newtheorem{definition}{Definition}

\theoremstyle{remark}

\title{Measurement of Trustworthiness of the Online Reviews}
\author{Corresponding Author:Dipankar Das \\Associate Professor of Economics, Finance and Data Science, Goa Institute of Management \\\texttt{dipankar3das@gmail.com, dipankar@gim.ac.in}
}
\date{}
\begin{document}
	\maketitle
	\begin{abstract}
		Online review platforms shape consumer decisions, yet reported ratings and comments may be unreliable when reviewers behave inconsistently. This paper models online reviews as a sequential choice problem and proposes a formal rationality pattern function that links a reviewer's current review to their revealed preference history. Building on a two-way consistency axiom for choices from nested sets, we derive an object-specific support trajectory $\boldsymbol\omega_x$ and an associated degree measure in $[0,1]$ (Average Propensity to Choose a Pattern, APCP) that quantifies review trustworthiness. The measure is designed to support information updating and reduce asymmetric information by discounting reviews that are inconsistent with past behavior. A worked example illustrates how the approach assigns trustworthiness grades to reviews for different objects and how these grades can complement aggregate rating statistics. Finally, a generalized theory has been established.
	\end{abstract}

	\textbf{JEL Classification Code:} D	
	\begin{center}
		\textbf{Measurement of the Trustworthiness of the Online Reviews}
	\end{center}
	\section{Introduction}
    Online reviews are now a ubiquitous layer of socio-technical information infrastructure: they represent experiential information in compact forms (e.g., star ratings and text), are organized and surfaced by platform systems (e.g., ranking, filtering, and recommendation), and shape how people interpret quality and make decisions at scale. However, a persistent information-quality problem remains: the observable content of a review rarely indicates whether the reviewer’s expressed evaluation is consistent with the reviewer's own prior choices and comparisons. Previous research has examined review impacts, deception and manipulation, helpfulness, and trust-building mechanisms in digital markets \citep{huang2006herding,chen2017should,filieri2016makes,racherla2012perceived,utz2012consumers,song2020effect,qin2022integrated,punetha2023bayesian}. What is still underdeveloped is a formal review-level measure of credibility that links a reviewer’s current report to that reviewer’s revealed preference history.

	\indent To address this gap, we conceptualize reviewing as a human--information interaction embedded in a sequential choice process. A buyer first navigates through the nested opportunity sets (screening, comparing and selecting) and later produces an evaluative information artifact (a rating and narrative) about the focal object. From an information-science perspective, a review is more trustworthy when the informational claim it conveys is compatible with the previously revealed choice pattern of the reviewer for that object. This approach connects online reviews with sequential choice theory \citep{rubinstein2006model} and to prediction combination, where noisier individual signals should receive less weight in aggregation \citep{bates1969combination,clemen1986combining,winkler1992sensitivity,winkler2004multiple,wang2023forecast}.

	\indent We contribute (i) an object-specific rationality pattern function that represents the compatibility between a review and the sequential behavior of the reviewer, and (ii) a normalized degree measure, the Average Propensity to Choose a Pattern (APCP) that maps this compatibility to a trustworthiness score $[0,1]$. Together, these constructs support a transparent and reproducible assessment of review credibility and enable trust-adjusted information indices, rankings, and summaries in settings where platforms can observe browsing, comparison, or purchase histories.

	\indent The remainder of the paper proceeds as follows. The next subsection situates the contribution in the literature on reviews, information quality, and trust. Section~2 introduces the sequential-choice framework, derives the object-specific rationality pattern, and defines APCP. The results section illustrates how the proposed measure differentiates between reviews that may appear similar on the surface but carry different degrees of trustworthiness.
	\subsection{Related Literature}
	Despite the explosive growth of electronic commerce, very little is known about how consumers make purchase decisions in such settings. During purchase decisions, consumers often cannot evaluate all available alternatives in great depth and, thus, tend to use two-stage processes to reach their decisions. In the first stage, consumers typically screen for many available products and identify a subset of the most promising alternatives\citep{haubl2000consumer}.This means that there is a sequential choice problem. Consumers have limited cognitive resources and may not be able to process the potentially vast amounts of information about these alternatives. Consumer trust plays an important role in online commerce. Then what makes customers return to an e-vendor? This question has been answered in \citep{gefen2003trust}. trust-building mechanisms have been explained to exist \citep{awad2008establishing}. Research on user trustworthiness on social networks is gaining attraction, with many interesting studies conducted in recent years. A recent paper has studied a systematic review of it and its future directions in \citep{alkhamees2021user}. Creating a trust-based relationship with customers is a primary benefit that is nearly as important as the technical attributes of the website, such as usefulness. Perfect competition is rarely seen in practice, where market imperfections mitigate such factors as perfect information, market prices, and features of commodities. Combined with the criteria for product selection, the multiple sources of information available through the e-commerce channel can reduce consumers' search costs and support their intelligence phase and, subsequently, they can lead to the development of a plan to evaluate the alternatives available for making the decision. Online support for gathering information leads to better development of criteria for evaluating decision alternatives. Recognizing the need for information gathering (intelligence phase), the channel owner can provide links to an independent comparison and rating website, thus supporting the consumer's decision-making process \citep{kohli2004understanding}. In this respect, review ratings and review comments play an important role \citep{zhang2023reviewfeatures}. However, a correlation between good reviews and high demand may be spurious, induced by an underlying correlation with unobservable quality signals \citep{reinstein2005influence}.The author has explained that an early positive review increases consumers' attention, and the influence effect differs between categories. The results suggest that expert reviews can be an important mechanism for transmitting information about goods of uncertain quality \citep{reinstein2005influence}. A theoretical analysis of the impact of this behavior on firm profits and consumer surplus has been done in \citep{dellarocas2006strategic}. The study shows that in most cases, all companies will manipulate their online ratings. This implies that consumers should expect a certain amount of hype to be present in most online forums and must learn to compensate for it (by properly deflating what they see and read in such forums) when making inferences from such information.The online marketplace should have a mechanism to eliminate the problem of the lemon market. Because numerical ratings do not convey much information beyond text comments, could feedback forum designers attempt to codify and summarize all seller text comments or enable buyers to report their past experiences in terms of meaningful and quantifiable categories \citep{pavlou2006nature}. Here we ask the question of how to identify the reviewer's past behavior. A study finds that the dispersion of the ratings is positively correlated with sales growth and that the mean of the high end of the set of ratings is positively correlated with growth \citep{clemons2006online}. In \citep{chen2008online} a normative model has been studied to address several important strategic issues related to consumer reviews. The impact of changes in ranking on product adoption and the lack of impact of product reviews on the adoption of popular products studied in \citep{duan2009informational}. An interesting study has been conducted, where it has been shown that online reviewers are competing to get attention as a scarce resource. This study tries to understand how online users, especially online reviewers, compete for scarce resources and attention when writing online reviews\citep{shen2013competing}. We still have a limited understanding of the individual's decision to contribute these opinions.The effects of selection and adjustment that influence the decision to evaluate have been studied here in\citep{moe2012online}.The type of product moderates the effect of the depth and degree of review on the usefulness of a review \citep{mudambi2010research,chua2015helpfulness}. One can learn from and be affected by other consumers. opinions and the actual purchase decisions of others. Opinions or word of mouth have been studied in \citep{chen2011online}. The social influence of online ratings of others has been studied in \citep{sridhar2012social,shen2016herd}. Consumers use reviewer disclosure of identity-descriptive information to supplement or replace product information when making purchase decisions and evaluating the usefulness of online reviews. The paper studies this in \citep{forman2008examining} and suggests that online retailers may be able to increase sales on their sites by taking actions to encourage reviewers to reveal more descriptive identity content about themselves. Understanding the relationship between promotional marketing of companies and word of mouth in the context of a third-party review platform has been studied in \citep{lu2013promotional}. Online reviews should create a producer and consumer surplus by improving consumers' ability to evaluate unobservable product quality. Fake or promotional online reviews are also available \citep{banerjee2017deceived}. Suboptimal choices and consumer mistrust of reviews are the two important impacts. These are due to Fake or promotional online reviews. A methodology for empirically detecting review manipulation has been given in \citep{mayzlin2014promotional}. However, no such methods are available to understand the trustworthiness of the reviewer. The present article tries to provide a method for indexing and identifying patterns in online reviewers.
	
	\section{Theory and Methods}
	\subsection{Theoretical Motivation}
	Let \(y\) denote the variable being forecast and suppose that \(K\) forecasts \(\hat y_1,\ldots,\hat y_K\) of \(y\) are available. We aim to combine them into an aggregate forecast. For example, an e-commerce customer may wish to forecast the quality of an item from three preceding predictions \(\hat y_1,\hat y_2,\hat y_3\). These predictions represent conditional mean quality ratings supplied by the first three reviewers; the next customer is indexed by \(t=4\).
	
	Let \(N_t\) be the number of reviews available through period \(t\), and let \(\bar q_t\) be their mean rating. A period-specific forecasting rule \(F_t\) can be written as
	
	\begin{equation}
		\hat y_t=F_t(N_{t-1},\bar q_{t-1}).
	\end{equation}
	
	Thus, the forecasts from the past three customers can be expressed as
	
	\begin{equation}
		\begin{aligned}
			\hat y_1 &= F_1(N_0,\bar q_0), \\
			\hat y_2 &= F_2(N_1,\bar q_1), \\
			\hat y_3 &= F_3(N_2,\bar q_2).
		\end{aligned}
	\end{equation}
	
	The combined forecast for the fourth potential customer is
	
	\begin{equation}
		\hat y_4=G(\hat y_1,\hat y_2,\hat y_3)=F_4(N_3,\bar q_3).
	\end{equation}
	
	This combined forecast is derived using Bayesian Information updating. A critical assumption here is that past reviewers are trustworthy and have adhered to a rational choice process in previous periods. If past reviewers (forecasters) are not trustworthy, the combined forecast may have a statistically minimum error with respect to the mean squared error (MSE) loss function but could still be biased or misleading.
	
	This paper proposes a method to identify and exclude less trustworthy reviews to improve the reliability of the combined forecast.
	
	\subsection{Statement of Intended Contribution}
\begin{figure}[H]
		\centering\includegraphics[width=\columnwidth]{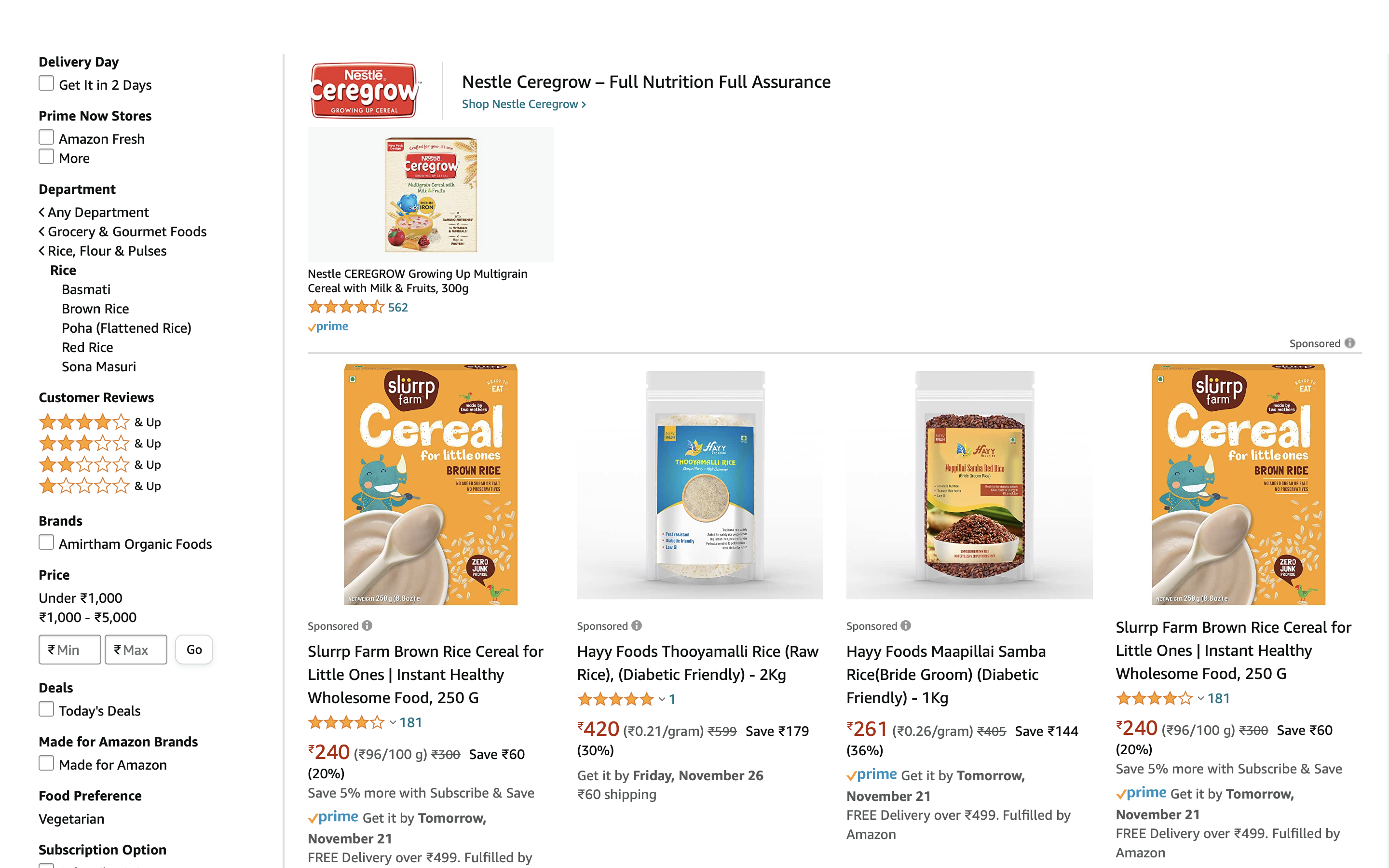}\caption{Review Rating-Example from amazon.in}\label{figure3}
	\end{figure}

\begin{figure}[H]
		\centering
		\includegraphics[width=0.45\textwidth]{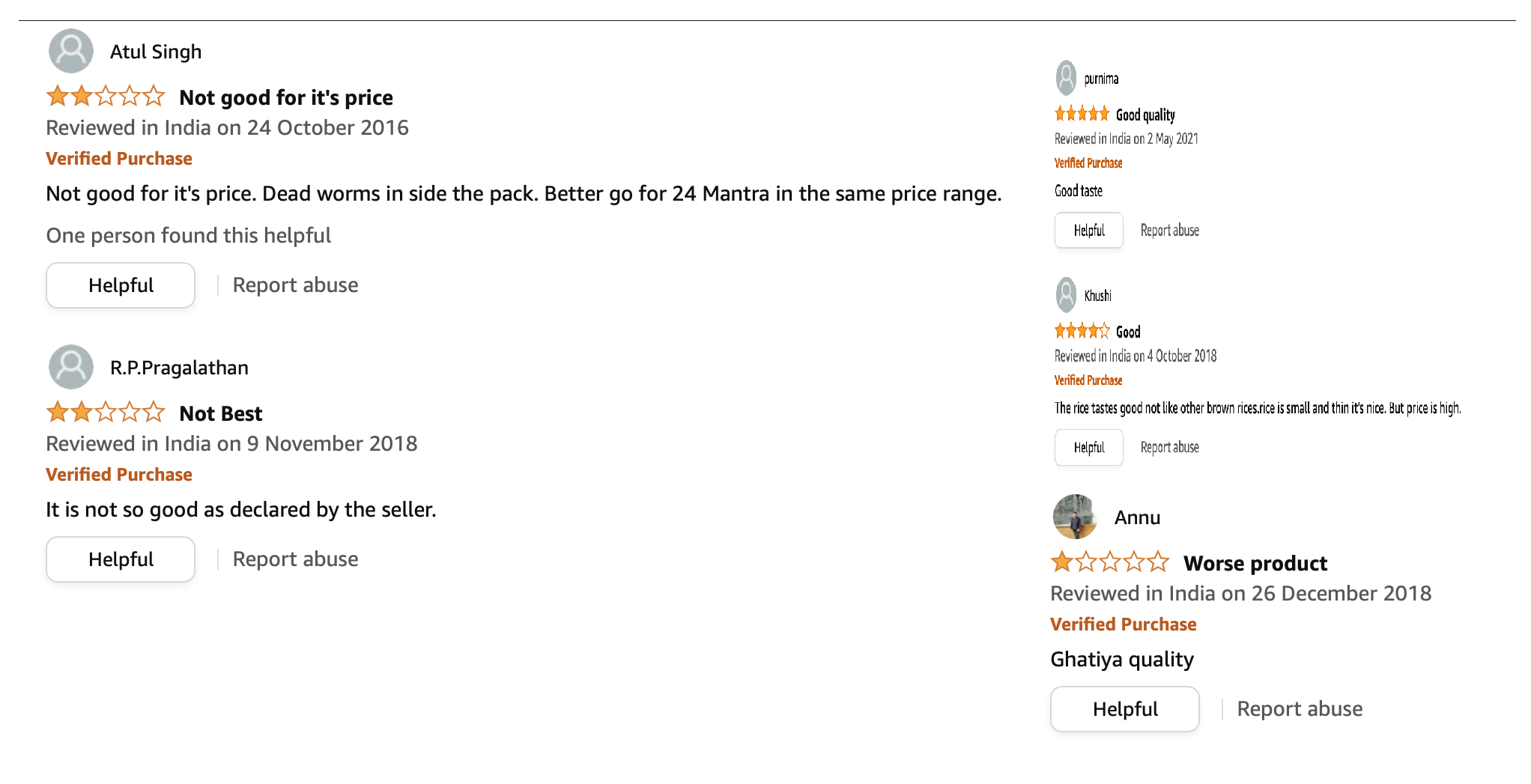}
		\caption{Review comments example from Amazon.in}
		\label{figure4}
	\end{figure}

	The contribution is developed through a simple two-period, four-object example that mirrors common online-shopping behavior. Figures \ref{figure3} and \ref{figure4} illustrate the kind of ratings and textual comments that are typically observed on digital platforms. The formal task is not to predict the average rating of a product, but to assess whether a specific review is credible given the reviewer’s own sequence of past choices.\\
	\textcolor{black}{Consider a decision-maker who chooses from the universe $\mathcal X=\{M,N,V,Z\}$ over two periods, $t=1,2$. Suppose the extracted strict preference orders are $M\succ_1 N\succ_1 V\succ_1 Z$ and $Z\succ_2 V\succ_2 N\succ_2 M$. These rankings can be inferred from platform records of sequential comparison and choice. The same reviewer then leaves a verbal review and star rating for each object in period $t=2$, as shown in Table~\ref{tab:table1}.}
	\begin{table}{}
		\begin{center}
			\caption{Review Comments and Ratings of the Given Decision Maker}
			\label{tab:table1}
			\begin{adjustbox}{max width=\columnwidth}
			\begin{tabular}{ |p{3.5cm}|p{3cm}|p{3cm}|p{3cm}|}
				\hline
				\multicolumn{4}{|c|}{Hypothetical Example}\\
				\hline
				Commodity/Object type & Review Comments & Review Ratings & Degree of Trustworthiness\\\hline
				M & Bad product & $ \circledast\bigcirc\bigcirc\bigcirc\bigcirc $ & \\
				N & Not so good product & $ \circledast\circledast\bigcirc\bigcirc\bigcirc $ & \\
				V & Relatively good product & $ \circledast\circledast\circledast\bigcirc\bigcirc $ & \\
				Z & Premium product & $ \circledast\circledast\circledast\circledast\circledast $ & \\\hline
				Or, Z & Not a Premium product & $ \circledast\circledast\circledast\circledast\bigcirc $ & \\
				\hline
			\end{tabular}
			\end{adjustbox}
		\end{center}
	\end{table}
	The empirical question is therefore to fill the last column of Table~\ref{tab:table1}: how trustworthy is each review, conditional on the previously revealed ordering of the reviewer? We answer this by attaching an object-specific pattern and a normalized degree measure to every review.
		\textcolor{black}{\begin{definition}
				\textbf{Average Propensity to Choose a Pattern (APCP).} Let $\boldsymbol\omega=(d_1,\ldots,d_T)\in\{0,1,\ldots,n-1\}^{T}$ be the revealed-support trajectory of an object across $T$ periods. For $n\geq2$ and $T\geq2$, define
				\[
				\nu_{n,T}(\boldsymbol\omega)
				=1-\frac{1}{(T-1)(n-1)}
				\sum_{t=1}^{T-1}\lvert d_{t+1}-d_t\rvert.
				\]
				Then $0\leq\nu_{n,T}(\boldsymbol\omega)\leq1$. Higher values indicate smaller average temporal discrepancy and hence greater behavioral continuity. For a single period, set $\nu_{n,1}=1$.
		\end{definition}}
	\paragraph{Note (how to read $\nu_{n,T}$).} For each period $t$, the support count $d_t$ is the outdegree of the object in the revealed strict-preference graph, i.e., the number of other objects it is strictly preferred to in that period (so $0\le d_t\le n-1$). The trajectory $\boldsymbol\omega=(d_1,\ldots,d_T)$ records how this support count evolves over time. The term $\sum_{t=1}^{T-1}|d_{t+1}-d_t|$ adds up the period-to-period changes in support. Dividing by $(T-1)(n-1)$ normalizes by the maximum possible total change, and subtracting from $1$ converts ``discrepancy'' into a continuity score: $\nu_{n,T}=1$ when the outdegree is constant across periods, and $\nu_{n,T}=0$ when the outdegree makes the largest possible jump in every transition.
	In this framework, a textual review and its star rating are not interpreted in isolation. They are evaluated against the reviewer’s path of revealed comparisons. The main contribution of the paper is therefore twofold: it derives an object-level rationality pattern from sequential choice and converts that pattern into a trustworthiness score that can be used for review weighting, ranking, and information aggregation.
		\subsection{Objective of the Study}
		The objective is to determine whether a reviewer's comment and star rating are behaviorally supported by that reviewer's revealed choices. For each object, the method extracts a support trajectory from the period-specific preference rankings and converts its temporal discrepancy into an APCP trustworthiness degree. A zero entry means that the object is not strictly preferred to any other observed object in that period. The resulting score supplements, rather than replaces, the observed comment and rating.
		\subsection{Model}
		Let $\mathcal X=\{x_1,\ldots,x_n\}$ be a finite universe of $n$ objects, and let $\mathfrak B=2^{\mathcal X}\setminus\{\varnothing\}$ be the family of nonempty choice problems. In period $t$, let $\succcurlyeq_t$ be a complete, reflexive, and transitive preference relation on $\mathcal X$. Its strict part is
		\begin{equation}
			x\succ_t y
			\quad\Longleftrightarrow\quad
			x\succcurlyeq_t y\ \land\ \neg(y\succcurlyeq_t x),
		\end{equation}
		which is asymmetric and transitive. A single-valued choice function is a map
		\begin{equation}
			C_t:\mathfrak B\longrightarrow\mathcal X,
			\qquad C_t(B)\in B\quad\text{for every }B\in\mathfrak B.
		\end{equation}
		For nested problems $B_1\subseteq B_2$, contraction consistency and expansion consistency are, respectively,
		\begin{align}
		&C_t(B_2)\in B_1
		\Longrightarrow C_t(B_1)=C_t(B_2),\label{eq:contraction-consistency}\\
		&C_t(B_1)\succcurlyeq_t y\ \text{for every }y\in B_2\setminus B_1
		\Longrightarrow C_t(B_2)=C_t(B_1).\label{eq:expansion-consistency}
		\end{align}
		\begin{definition}[Strong and weak choice consistency]
			Following the terminology used in this paper, $C_t$ is \emph{strongly consistent} when it satisfies Eq.~\eqref{eq:contraction-consistency} and \emph{weakly consistent} when it satisfies Eq.~\eqref{eq:expansion-consistency}.
		\end{definition}
		A choice function is \emph{two-way consistent} when it satisfies both conditions.
		\begin{theorem}\label{14}
			Let $C_t$ be two-way consistent and let $B_1\subseteq B_2$. The choice is preserved under contraction whenever $C_t(B_2)\in B_1$, and it is preserved under expansion whenever $C_t(B_1)$ is weakly preferred to every alternative newly introduced in $B_2\setminus B_1$.
		\end{theorem}
		\begin{proof}
			The first conclusion follows directly from Eq.~\eqref{eq:contraction-consistency}; the second follows directly from Eq.~\eqref{eq:expansion-consistency}.
		\end{proof}
		\paragraph{Interpretation.} Contraction consistency requires the larger-set choice to remain feasible after contraction. Expansion consistency requires the smaller-set choice to dominate the newly introduced alternatives. These conditions provide the stable-choice benchmark used in the empirical pattern analysis below.
	\subsection{Statistical Implication}
\begin{figure}[H]
		\centering\includegraphics[scale=0.40]{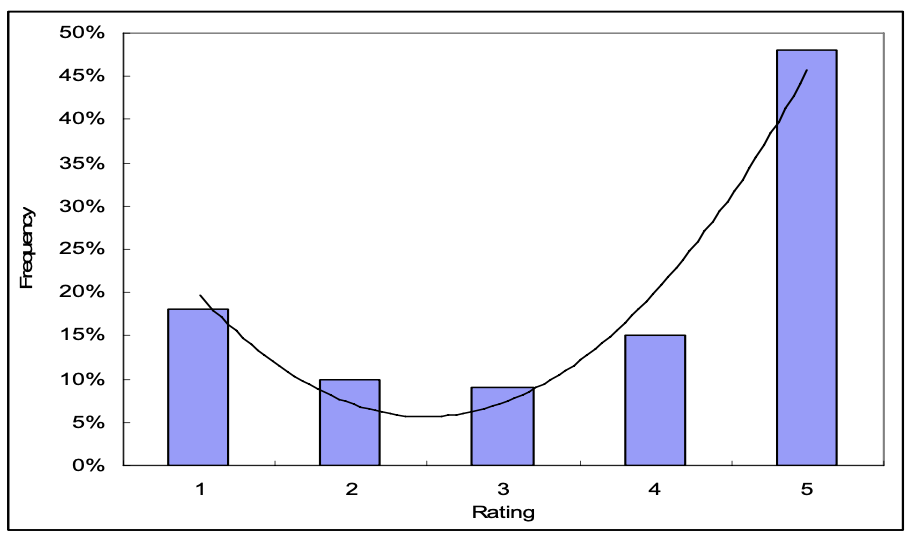}\caption{\textcolor{black}{Distribution of the Ratings on Amazon.com (A U-shaped curve)}}\label{figure5}
	\end{figure}

	Theorem~\ref{14} states the conditions under which a decision-maker preserves a reference choice as the feasible set contracts or expands. A study shows that $53\%$ of the products have a bimodal and non-normal distribution. The average score for these products does not necessarily reveal the product's quality and may provide misleading recommendations. Figure~\ref{figure5} shows a U-shaped distribution of ratings for a music CD on Amazon.com.
	However, when they asked an unbiased population of $66$ students to test the product's review distribution, they obtained a distribution close to a Gaussian \citep{hu2006can}. This contrast motivates a reviewer-level consistency measure rather than reliance on the mean rating alone. Theorem~\ref{14} supplies the behavioral stability benchmark; it does not, by itself, correct or identify a particular rating distribution.
	\subsection{The Choice Process}
	The platform records a sequence of nested filtering and selection stages. This permits the reviewer’s final report to be compared with the reviewer’s own revealed choices.
	\begin{definition}[Object]
		An object $x_i$ is represented by a vector of $d$ discrete attributes. Thus $x_i\in\mathbb Z_{+}^{d}$, and the finite product universe is
		\begin{equation}
			\mathcal X=\{x_1,\ldots,x_n\}\subseteq\mathbb Z_{+}^{d}.
		\end{equation}
	\end{definition}
	The integer $d$ denotes the number of attributes, whereas $n=\lvert\mathcal X\rvert$ denotes the number of objects. The ambient attribute space is discrete and is not assumed to be convex. Finiteness follows from the finite platform catalog, not from discreteness alone.

	\paragraph{Stage 1: attainable set.}
	Let $B_t$ be a constraint matrix and $c_t$ a constraint vector. The attainable set in period $t$ is
	\begin{equation}\label{key2b}
		F_t=\{x\in\mathcal X:B_t x\leq c_t\}.
	\end{equation}
	This set represents the objects satisfying the reviewer’s budget and attribute requirements.

	\paragraph{Stage 2: wish list.}
	Let $\Gamma(F_t)$ denote the set of finite ordered lists whose entries belong to $F_t$. A single-valued list choice has the form
	\begin{equation}
		D_t:\Gamma(F_t)\longrightarrow F_t,
		\qquad D_t(L)\in L.
	\end{equation}
	A wish-list selector may instead be set-valued. We write
	\begin{equation}
		\Phi_W:\Gamma(F_t)\longrightarrow 2^{F_t}\setminus\{\varnothing\},
		\qquad W_t=\Phi_W(L_t)\subseteq F_t.
	\end{equation}

	\paragraph{Stage 3: cart.}
	The cart selector is
	\begin{equation}
		\Phi_A:2^{F_t}\setminus\{\varnothing\}
		\longrightarrow 2^{F_t}\setminus\{\varnothing\},
		\qquad A_t=\Phi_A(W_t)\subseteq W_t.
	\end{equation}

	\paragraph{Stage 4: final choice.}
	Under the single-purchase assumption, the final selected object is
	\begin{equation}
		x_t^{*}=C_t(A_t)\in A_t,
		\qquad S_t=\{x_t^{*}\}\subseteq A_t\subseteq W_t\subseteq F_t.
	\end{equation}
	A bundle-valued extension would replace the singleton $S_t$ by a nonempty subset of $A_t$.

	\begin{definition}[Rational choice]
		When $\succ_t$ is a strict total order, the decision-maker chooses the $\succ_t$-best feasible object from every recorded choice problem.
	\end{definition}
	The stable-choice benchmark for the nested stages is
	\begin{equation}
		C_t(S_t)=C_t(A_t)=C_t(W_t)=C_t(F_t)=x_t^{*}.
	\end{equation}
	Observed departures from this equality across recorded transactions motivate the object-specific trajectory developed below.
	\subsection{Single Agent and Two Periods Model of the Rationality Pattern}
	This section applies the sequential-choice framework to one decision-maker, the four-object universe $\mathcal X=\{M,N,V,Z\}$, and two periods. The strict preference orders are
	\begin{equation}
		M\succ_1 N\succ_1 V\succ_1 Z,
		\qquad
		Z\succ_2 V\succ_2 N\succ_2 M.
	\end{equation}
	The reviewer supplies the period-2 comments and star ratings shown in Table~\ref{tab:table1}. We first represent each period separately and then form object-specific two-period support trajectories.
	\subsection{Algorithm for Degree of Rationality Analysis}
		\paragraph{Rationality Analysis Steps.}
		\begin{enumerate}[label=\textbf{STEP-(\Roman*)}, leftmargin=*]
			\item Initialize the object universe $\mathcal X$.
			\item Define preferences for Periods 1 and 2.
			\item Construct the period-specific preference matrices.
			\item Calculate each object’s revealed-support count.
			\item Form the ordered support trajectories.
			\item Group two-period trajectories by signed displacement.
			\item Compute the APCP trustworthiness degree.
		\end{enumerate}
	\begin{center}
		\textbf{STEP-(I)}
	\end{center}
	\subsubsection{Preference in Period \texorpdfstring{$ t=1$}{t=1} and Identification of the Pattern}
	The recorded nested sets in period 1 are
	\begin{equation}
		S_1=\{M\}\subseteq A_1=\{M,N\}
		\subseteq W_1=\{M,N,V\}
		\subseteq F_1=\mathcal X.
	\end{equation}
	They reveal the strict order $M\succ_1 N\succ_1 V\succ_1 Z$, so the selected object is $x_1^*=M$. For the fixed object order $(M,N,V,Z)$, define the preference matrix $U_t=(u_{ij}^{(t)})$ by $u_{ij}^{(t)}=1$ if $x_i\succ_t x_j$ and $u_{ij}^{(t)}=0$ otherwise. The period-1 matrix is
	\begin{center}
		$U_1=
		\begin{array}{c c c c } &
			\begin{array}{c c c c} M & N & V & Z \\
			\end{array}
			\\
			\begin{array}{c c c c}
				M \\
				N\\
				V\\
				Z\\
			\end{array}
			&
			\left[
			\begin{array}{c c c c}
				0 & 1 & 1 & 1 \\
				0 & 0 & 1 & 1 \\
				0 & 0 & 0 & 1 \\
				0 & 0 & 0 & 0
			\end{array}
			\right]
		\end{array}
		$
	\end{center}
	The revealed-support count of object $x$ is the outdegree
	\begin{equation}\label{eq:support-count-preliminary}
		d_t(x)=\left|\{y\in\mathcal X:x\succ_t y\}\right|.
	\end{equation}
	Thus $(d_1(M),d_1(N),d_1(V),d_1(Z))=(3,2,1,0)$. The row-major binary encoding is
	\begin{equation}
		\mathbf p_1=\operatorname{vec}_{\mathrm r}(U_1)
		=(0,1,1,1;0,0,1,1;0,0,0,1;0,0,0,0).
	\end{equation}
	Equivalently, the nonzero row-support words are summarized by
	\begin{equation}
		\mathbf r_1=(111,11,1,\epsilon),
	\end{equation}
	where $\epsilon$ denotes the empty binary word. The word lengths are precisely the support counts $(3,2,1,0)$.
	The associated preference graph $G_1$ is shown in Figure~\ref{figure1}; it is acyclic.
	\begin{figure}[t]
		\centering
		\begin{tikzpicture}[node distance=1.05cm, every node/.style={circle,draw,minimum size=6mm}, >={Latex}]
			\node (M) {M};
			\node[right=of M] (N) {N};
			\node[right=of N] (V) {V};
			\node[right=of V] (Z) {Z};
			\draw[->] (M) -- (N);
			\draw[->] (N) -- (V);
			\draw[->] (V) -- (Z);
			\draw[->,bend left=35] (M) to (V);
			\draw[->,bend left=45] (M) to (Z);
			\draw[->,bend left=35] (N) to (Z);
		\end{tikzpicture}
		\caption{Preference graph}\label{figure1}
	\end{figure}  
	More generally, a sequence of transactions produces matrices $\{U_t\}_{t=1}^{T}$ and graphs $\{G_t\}_{t=1}^{T}$. The same support words may be read in decreasing set-size order as $(111,11,1)$ or in increasing set-size order as $(1,11,111)$; these are ordered sequences rather than sets.
	\paragraph{Numerical calculation (APCP from outdegrees).} From $U_1$ the row sums give the period-1 support counts $(d_1(M),d_1(N),d_1(V),d_1(Z))=(3,2,1,0)$. In Step-(II) we compute from $U_2$ that $(d_2(M),d_2(N),d_2(V),d_2(Z))=(0,1,2,3)$. Hence for $n=4$ objects and $T=2$ periods the revealed-support trajectories are $\boldsymbol\omega_M=(3,0)$, $\boldsymbol\omega_N=(2,1)$, $\boldsymbol\omega_V=(1,2)$, and $\boldsymbol\omega_Z=(0,3)$. Applying Definition~(APCP),
	\[
	\nu_{4,2}(\boldsymbol\omega_x)
	=1-\frac{1}{(2-1)(4-1)}\,\bigl|d_2(x)-d_1(x)\bigr|
	=1-\frac{1}{3}\,\bigl|d_2(x)-d_1(x)\bigr|.
	\]
	For example, for $M$ we have $|d_2(M)-d_1(M)|=|0-3|=3$, so $\nu_{4,2}(\boldsymbol\omega_M)=1-3/3=0$. For $N$ we have $|1-2|=1$, so $\nu_{4,2}(\boldsymbol\omega_N)=1-1/3=2/3$ (and similarly $\nu_{4,2}(\boldsymbol\omega_V)=2/3$, $\nu_{4,2}(\boldsymbol\omega_Z)=0$).
	\paragraph{Remarks 4} 
	\textcolor{black}{When an agent starts selecting the alternative/element/object from the list, the agent tries to match the pre-determined attributes/messages with the actual messages each object carries. Those messages' degree of fulfillment, degree of non-fulfillment, and degree of indeterminacy play an important role in selecting the element from the given list. Say the set of messages to buy a cell phone are (i) product information, (ii) customers' reviews, (iii) after-sale services, etc. Suppose the agent gets perfect knowledge/messages/information about each attribute. In that case, the product with the highest information will be used as a satisfactory threshold, and the search will stop either to it or near it. Selecting alternatives will be easier if the alternatives carry this information. The agent tries to minimize the degree of indeterminacy and select the alternative that carries the highest information. The search process will continue until that alternative carries the highest information.}\\
	\indent For a rational choice, contraction and expansion through the nested stages must select the same object. In the binary matrix, $1$ denotes a strict preference and $0$ its absence. Changes in an object’s support count across periods are the inputs to the consistency measure.
	\begin{definition}\textbf{Information Index.}
		Let $H:\mathcal X\to\mathbb{R}$ be an index assigned to each alternative. For any choice problem $B\in\mathfrak B$, define its information index by
		\[
			\mathcal H_t(B)=H\bigl(C_t(B)\bigr).
		\]
	\end{definition}
	\begin{center}
		\textbf{STEP-(II)}
	\end{center}
	\subsubsection{Preference in Period \texorpdfstring{$ t=2 $}{t=2} and Identification of the Pattern}
	The period-2 order is $Z\succ_2 V\succ_2 N\succ_2 M$, so the selected object is $x_2^*=Z$. Its preference matrix is
	\begin{center}
		$U_2=
		\begin{array}{c c c c } &
			\begin{array}{c c c c} M & N & V & Z \\
			\end{array}
			\\
			\begin{array}{c c c c}
				M \\
				N\\
				V\\
				Z\\
			\end{array}
			&
			\left[
			\begin{array}{c c c c}
				0 & 0 & 0 & 0 \\
				1 & 0 & 0 & 0 \\
				1 & 1 & 0 & 0 \\
				1 & 1 & 1 & 0
			\end{array}
			\right]
		\end{array}
		$
	\end{center}
	The support-count vector is $(d_2(M),d_2(N),d_2(V),d_2(Z))=(0,1,2,3)$, and
	\begin{align}
		\mathbf p_2&=\operatorname{vec}_{\mathrm r}(U_2)
		=(0,0,0,0;1,0,0,0;1,1,0,0;1,1,1,0),\\
		\mathbf r_2&=(\epsilon,1,11,111).
	\end{align}
	\paragraph{Remarks 5} The word-length multiset is the same in each period, but its assignment to objects is reversed. The ordered object-level trajectories therefore differ: for example, $N$ has the word trajectory $(11,1)$, whereas $V$ has $(1,11)$. Parentheses are essential because temporal order matters.
	\begin{center}
		\textbf{STEP-(III)}
	\end{center}
	\subsubsection{Joint Preference for Periods 1 and 2}
	The period-preserving joint encoding is the concatenation
	\begin{equation}
		\mathbf p_{1,2}=(\mathbf p_1,\mathbf p_2)\in\{0,1\}^{32}.
	\end{equation}
	For comparison, define the elementwise Boolean union $\bar U=U_1\lor U_2$. This aggregate matrix is
	\begin{center}
		$\bar U=
		\begin{array}{c c c c } &
			\begin{array}{c c c c} M & N & V & Z \\
			\end{array}
			\\
			\begin{array}{c c c c}
				M \\
				N\\
				V\\
				Z\\
			\end{array}
			&
			\left[
			\begin{array}{c c c c}
				0 & 1 & 1 & 1 \\
				1 & 0 & 1 & 1 \\
				1 & 1 & 0 & 1 \\
				1 & 1 & 1 & 0
			\end{array}
			\right]
		\end{array}
		$
	\end{center}
	\begin{center}
		\begin{figure}[t]
			\centering
			\begin{tikzpicture}[every node/.style={circle,draw,minimum size=6mm}, >={Latex}]
				\node (M) at (0,1.4) {M};
				\node (N) at (1.8,1.4) {N};
				\node (V) at (0,0) {V};
				\node (Z) at (1.8,0) {Z};
				\draw[<->] (M) -- (N);
				\draw[<->] (M) -- (V);
				\draw[<->] (M) -- (Z);
				\draw[<->] (N) -- (V);
				\draw[<->] (N) -- (Z);
				\draw[<->] (V) -- (Z);
			\end{tikzpicture}
			\caption{Boolean-union preference graph $\bar G$}\label{figure2}
		\end{figure} 
	\end{center} 	
	Every vertex of $\bar G$ has outdegree $3$, and every unordered pair forms a two-cycle. Consequently, $\bar U$ loses the temporal direction of the comparisons. The concatenated vector $\mathbf p_{1,2}$, or equivalently the object-specific trajectories defined below, must be retained for dynamic analysis.
	\paragraph{Remarks 6} The matrices $U_1$ and $U_2$ represent period-specific preferences; $\bar U$ represents only their Boolean union. These objects are mathematically distinct and are not used interchangeably.
	\begin{center}
		\textbf{STEP-(IV)}
	\end{center}
	\begin{definition}[Stop and run]
		In a row of $U_t$, a zero records the absence of a strict preference and a one records a revealed strict preference. The diagonal zero identifies the focal object; the row sum, rather than the physical position of the ones, is its revealed-support count. A binary word of length $d_t(x)$ may be used as a shorthand run, with $\epsilon$ denoting the empty word when $d_t(x)=0$.
	\end{definition}
	\paragraph{Numerical Calculation 1} Reading the row sums of $U_1$ and $U_2$ gives
	\begin{center}
		$ 
		\begin{bmatrix}
			M & N & V & Z & M & N & V & Z \\
			3 & 2 & 1 & 0 & 0 & 1 & 2 & 3\\
		\end{bmatrix}
		$
	\end{center}
	Each entry is the number of alternatives ranked below the corresponding object. These period-specific counts are the inputs to the rationality-pattern function.
	\begin{lemma}\label{lem:example-not-completely-irrational}
		Under the paper’s run-based terminology, the realized joint encoding is not completely unsupported.
	\end{lemma}
	\begin{proof}
		The period-1 row for $M$ contains three positive comparisons, represented by the word $111$. Hence at least one object has positive revealed support in the observed encoding.
	\end{proof}
	\begin{proposition}\label{prop:complete-reversal}
		The two period-specific rankings in the example are complete reversals. If pairwise agreement is scored as $+1$ for concordance and $-1$ for discordance, the aggregate score over the $\binom{4}{2}=6$ unordered pairs is $-6$; its normalized value is $-1$.
		\end{proposition}
		\begin{proof}
			Period 1 ranks $M\succ_1N\succ_1V\succ_1Z$, whereas period 2 ranks $Z\succ_2V\succ_2N\succ_2M$. Every unordered pair is reversed, so all six pairwise scores equal $-1$. Their sum is $-6$, and division by $\binom{4}{2}=6$ gives $-1$.
		\end{proof}
		\paragraph{Remark.} This aggregate reversal score describes the reviewer across all objects. It is distinct from the object-specific APCP degrees developed below.
	\subsection{The Rationality Pattern Function and the Measurement}
	This section derives a pattern for each particular object. Binary words provide a compact representation of within-period comparisons, whereas numeric trajectories record support across periods. The two representations are kept distinct.
	\begin{proposition}
		Let $\Sigma=\{0,1\}$ be the binary alphabet, let $\Sigma^m$ be the set of words of length $m$, and define
		\begin{equation}
			\Sigma^*=\bigcup_{m\geq0}\Sigma^m,
			\qquad
			\Sigma^+=\bigcup_{m\geq1}\Sigma^m,
			\qquad
			\Sigma^0=\{\epsilon\}.
		\end{equation}
		For a universe of $n$ objects, the positive run language is
		\begin{equation}
			L_R^{(n)}=\{1^q:1\leq q\leq n-1\}\subseteq\Sigma^+.
		\end{equation}
		Here $1^q$ records that the focal object is strictly preferred to $q$ other objects, and $\epsilon$ records an empty run.
	\end{proposition}
	\begin{proof}
		A row of a strict-preference matrix contains one binary entry for each possible comparison. A focal object with $q$ positive comparisons has a support word consisting of $q$ ones. Because $0\leq q\leq n-1$, its word is either $\epsilon$ or an element of $L_R^{(n)}$. These definitions follow the standard notation for formal languages \citep{hopcroft2001introduction}.
	\end{proof}
	\textcolor{black}{\paragraph{Remarks 9} The formal-language notation is used only for binary comparison words. APCP is evaluated on the ordered numeric trajectory of their lengths, not on membership in $\Sigma^+$ alone.}
	\paragraph{Numerical Calculation 2} For $\mathcal X=\{M,N,V,Z\}$, the period-ordered word trajectories are
	\begin{equation}
		\begin{aligned}
		\boldsymbol\rho_M&=(111,\epsilon),&
		\boldsymbol\rho_N&=(11,1),\\
		\boldsymbol\rho_V&=(1,11),&
		\boldsymbol\rho_Z&=(\epsilon,111).
		\end{aligned}
	\end{equation}
	\paragraph{Remarks 10} The two rankings select different leading objects: $M$ at $t=1$ and $Z$ at $t=2$. In particular, $(1,11)\neq(11,1)$ because these are ordered trajectories. Their numeric length trajectories are $(1,2)$ and $(2,1)$, respectively.
	\begin{lemma}\label{lem:strong-weak-patterns}
		For nested choice stages containing two and three alternatives, expansion consistency produces the ordered support-word sequence $(1,11)$ when the same object remains selected, whereas contraction consistency produces the reverse sequence $(11,1)$.
	\end{lemma}
	\begin{proof}
		Under expansion, a selected object that dominates the newly added alternative has one positive comparison in the pair and two in the triple. Under contraction, the same counts are encountered in reverse order.
	\end{proof}
	\paragraph{Direction and magnitude.} The trajectory type distinguishes constant, increasing, decreasing, and irregular paths. APCP depends on the magnitude of adjacent changes and is direction-neutral. Hence the ordered trajectories for $N$ and $V$ remain different even though their APCP degrees are equal.
	\subsubsection{Rationality Pattern Function}
	\begin{definition}
		For a fixed universe of $n$ objects, let
		\begin{equation}
			\mathcal D_n=\{0,1,\ldots,n-1\}.
		\end{equation}
		For $T$ periods, the rationality outcome set is the Cartesian product
		\begin{equation}
			\tau_{n,T}=\mathcal D_n^T.
		\end{equation}
		Thus an outcome retains the period order and contains one support entry for each period.
	\end{definition}
	\paragraph{Numerical Calculation 3}
	When $T=2$ and $n=4$, $\mathcal D_4=\{0,1,2,3\}$ and $\tau_{4,2}=\mathcal D_4\times\mathcal D_4$ contains $4^2=16$ ordered trajectories:
	\begin{multline*}
		\tau_{4,2}=\{(0,0),(0,1),(0,2),(0,3),
		(1,0),(1,1),(1,2),(1,3),\\
		(2,0),(2,1),(2,2),(2,3),
		(3,0),(3,1),(3,2),(3,3)\}.
	\end{multline*}
	\begin{definition}For period-specific strict preferences $\succ_t$, define
		\begin{equation}
			d_t(x)=\left|\{y\in\mathcal X:x\succ_t y\}\right|.
		\end{equation}
		The object-specific rationality-pattern function is
		\begin{equation}
			\boldsymbol\omega:\mathcal X\longrightarrow\tau_{n,T},
			\qquad
			\boldsymbol\omega(x)=(d_1(x),\ldots,d_T(x)).
		\end{equation}
		For brevity, write $\boldsymbol\omega_x=\boldsymbol\omega(x)$.
	\end{definition}
	\begin{theorem}\label{thm:omega-membership}
		For every object $x\in\mathcal X$, the object-specific rationality pattern belongs to the outcome set:
		\begin{equation}
			\boldsymbol\omega_x\in\tau_{n,T}.
		\end{equation}
	\end{theorem}
	\begin{proof}
		In every period, $d_t(x)$ is an integer between $0$ and $n-1$. Hence every component of $\boldsymbol\omega_x$ belongs to $\mathcal D_n$, so $\boldsymbol\omega_x\in\mathcal D_n^T=\tau_{n,T}$.
	\end{proof}
	\paragraph{Numerical Calculation 4}
	When $T=2$, $n=4$, and $\mathcal X=\{M,N,V,Z\}$, the support table is
	\begin{equation}
		\begin{array}{c|cccc}
		& M&N&V&Z\\ \hline
		t=1&3&2&1&0\\
		t=2&0&1&2&3
		\end{array}.
	\end{equation}
	Therefore,
	\begin{equation}
		\boldsymbol\omega_M=(3,0),\quad
		\boldsymbol\omega_N=(2,1),\quad
		\boldsymbol\omega_V=(1,2),\quad
		\boldsymbol\omega_Z=(0,3).
	\end{equation}
	Each value belongs to $\tau_{4,2}$.
	\subsubsection{Exclusion and Inclusion of Objects in Different Periods}
	The value $0$ means that an observed object is bottom-ranked or otherwise has no positive comparison. If an object is absent from the observed choice problem, use a distinct missing-state symbol $\bot$. An application with varying availability may therefore use the extended state space $\mathcal D_n^{\bot}=\mathcal D_n\cup\{\bot\}$; $0$ and $\bot$ must not be conflated.
	\subsubsection{Degree of the Rationality Pattern} 
	The degree is object-specific: a review of object $x$ is paired with $\boldsymbol\omega_x$, not with the reviewer’s aggregate graph. The APCP value measures continuity of revealed support and is independent of whether the wording is positive or negative. Semantic agreement between the current comment and the current-period ranking can be assessed separately.
	\begin{definition}
		For $n\geq2$ and $T\geq2$, the APCP map is
		\begin{equation}
			\nu_{n,T}:\tau_{n,T}\longrightarrow[0,1],
		\end{equation}
		with
		\begin{equation}\label{eq:apcp-general}
			\nu_{n,T}(d_1,\ldots,d_T)
			=1-\frac{\sum_{t=1}^{T-1}\lvert d_{t+1}-d_t\rvert}
			{(T-1)(n-1)}.
		\end{equation}
	\end{definition}
	\subsubsection{Calculation of \texorpdfstring{$\nu_{n,2}(\boldsymbol\omega_x)$}{APCP}}
		For a two-period trajectory $p=(a,b)\in\tau_{n,2}$, define its signed displacement and modal difference by
		\begin{equation}\label{eq:modal-difference}
			\delta(p)=b-a,
			\qquad
			\Delta(p)=\lvert\delta(p)\rvert.
		\end{equation}
		For any fixed signed displacement $\delta\in\{-(n-1),\ldots,n-1\}$, the number of trajectories having that displacement is
		\begin{equation}
			m_n(\delta)=n-\lvert\delta\rvert.
		\end{equation}
		Thus $m_n$ is a combinatorial multiplicity, not a probability distribution. For two periods, Eq.~\eqref{eq:apcp-general} is equivalently
		\begin{equation}\label{eq:apcp-direct}
			\nu_{n,2}(p)
			=\frac{m_n(\delta(p))-1}{n-1}
			=1-\frac{\Delta(p)}{n-1}.
		\end{equation}
		\begin{corollary}\label{cor:apcp-unit-interval}
			For every $\boldsymbol\omega\in\tau_{n,T}$,
			\begin{equation}
				0\leq\nu_{n,T}(\boldsymbol\omega)\leq1.
			\end{equation}
		\end{corollary}
		\begin{proof}
			Each adjacent difference is between $0$ and $n-1$. Hence their sum is between $0$ and $(T-1)(n-1)$, and Eq.~\eqref{eq:apcp-general} gives the stated bounds.
		\end{proof}
		For the example, the modal differences of $\boldsymbol\omega_M,\boldsymbol\omega_N,\boldsymbol\omega_V,\boldsymbol\omega_Z$ are $3,1,1,3$, respectively.
	\subsection{Rationality Outcome Space and Trajectory Types}
	For a fixed universe of $n$ objects observed over $T$ periods, the outcome set is $\tau_{n,T}=\mathcal D_n^T$, and $\nu_{n,T}(\boldsymbol\omega_x)\in[0,1]$. Degree one is attained by a constant trajectory. Degree zero is attained when every adjacent change has the maximum magnitude $n-1$.
	\subsubsection{Trajectory Classification}
	Each element $\mathbf d=(d_1,\ldots,d_T)$ of $\tau_{n,T}$ is an ordered support trajectory. Its direction may be classified descriptively as
	\begin{enumerate}[label=(\roman*)]
		\item constant if $d_s=d_{s+1}$ for every adjacent pair;
		\item strictly increasing if $d_s<d_{s+1}$ for every adjacent pair;
		\item strictly decreasing if $d_s>d_{s+1}$ for every adjacent pair;
		\item irregular otherwise.
	\end{enumerate}
	This classification is not an APCP ranking: increasing and decreasing trajectories with the same absolute changes receive the same APCP degree.
	\begin{center}
		\textbf{STEP-(V)}
	\end{center}
	\subsubsection{Two-Period Displacement Table}
	For $T=2$, the number of strictly decreasing trajectories is $\binom n2$, and the number of constant or increasing trajectories is $\binom{n+1}{2}$. Therefore,
	\begin{equation}
		\binom n2+\binom{n+1}{2}=n^2=\lvert\tau_{n,2}\rvert.
	\end{equation}
	For $n=4$, the sixteen trajectories grouped by signed displacement are
	\begin{equation}\label{eq:displacement-table}
	\begin{array}{c|c|l}
	\text{Bar}&\delta=b-a&\text{Trajectories }(a,b)\\ \hline
	C&-3&(3,0)\\
	B&-2&(2,0),(3,1)\\
	A&-1&(1,0),(2,1),(3,2)\\
	D& 0&(0,0),(1,1),(2,2),(3,3)\\
	E& 1&(0,1),(1,2),(2,3)\\
	F& 2&(0,2),(1,3)\\
	G& 3&(0,3)
	\end{array}.
	\end{equation}
	For $T\geq3$, the full outcome space contains $n^T$ trajectories, including irregular paths. Equation~\eqref{eq:displacement-table} is a two-period grouping and is not asserted as a general ranking table.
	\begin{center}
		\textbf{STEP-(VI)}
	\end{center}
	\subsubsection{Signed-Displacement Multiplicity}
	The displacement groups produce the symmetric multiplicity profile $[1,2,3,4,3,2,1]$. This is a combinatorial count, not an assumed probability distribution.
	\begin{center}
		\begin{tikzpicture}
			\draw [->] (1,0) -- (9,0);
			\draw [->] (1,0)--(1,5);
			\node at (5,1) {$ (0,0) $};
			\node at (5,2) {$ (1,1 )$};
			\node at (5,3) {$ (2,2 )$};
			\node at (5,4) {$ (3,3) $};
			\node at (5,-1) {D};
			\node at (4,1) {$ (3,2) $};
			\node at (4,2) {$ (2,1 )$};
			\node at (4,3) {$ (1,0)$};
			\node at (4,-1) {A};
			\node at (6,1) {$ (0,1) $};
			\node at (6,2) {$ (1,2 )$};
			\node at (6,3) {$ (2,3 )$};
			\node at (6,-1) {E};
			\node at (7,1) {$ (0,2) $};
			\node at (7,2) {$ (1,3 )$};
			\node at (7,-1) {F};
			\node at (3,1) {$ (3,1) $};
			\node at (3,2) {$ (2,0)$};
			\node at (3,-1) {B};
			\node at (2,1) {$ (3,0) $};
			\node at (2,-1) {C};
			\node at (8,1) {$ (0,3) $};
			\node at (8,-1) {G};
		\end{tikzpicture}
	\end{center}
	\paragraph{Remarks 19} The first and second entries of $(a,b)$ record support in periods 1 and 2. Bars on opposite sides of $D$ have opposite signed displacements but the same modal difference. Therefore they receive the same APCP degree.
	The multiplicity assigned to each signed-displacement bar is shown in Table~\ref{tab:table2}.
	\begin{table}[H]
		\begin{center}
			\caption{Signed-displacement multiplicity profile}
			\label{tab:table2}
			\begin{tabular}{|c|c|c|}
				\hline
				\textbf{Bar} & \textbf{Displacement $\delta$} & \textbf{Multiplicity $m_4$} \\
				\hline
				C & $-3$ & 1 \\
				B & $-2$ & 2 \\
				A & $-1$ & 3 \\
				D & $0$ & 4 \\
				E & $1$ & 3 \\
				F & $2$ & 2 \\
				G & $3$ & 1 \\
				\hline
			\end{tabular}
		\end{center}
	\end{table} 
	\begin{center}
		\textbf{STEP-(VII)}
	\end{center}
	\subsubsection{APCP Function}
	For $n=4$ and $T=2$, the APCP normalization is
	\begin{equation}
		\nu_{4,2}(p)=\frac{m_4(\delta(p))-1}{3}
		=1-\frac{\lvert\delta(p)\rvert}{3}.
	\end{equation}
	For a bar $H\in\{A,B,C,D,E,F,G\}$, let $\nu_H$ denote the common APCP degree of the trajectories assigned to $H$. Therefore,
	\begin{equation}
		\bigl[\nu_C,\nu_B,\nu_A,\nu_D,\nu_E,\nu_F,\nu_G\bigr]
		=\left[0,\frac{1}{3},\frac{2}{3},1,\frac{2}{3},\frac{1}{3},0\right].
	\end{equation}
	These values are nonnegative, at most one, and symmetric about the zero-discrepancy bar $D$.
	\section{Results}
	This section applies the object-specific trajectories and APCP normalization to the illustrative two-period, four-object example.
	
	\subsection{Analysis of Object-Specific Trustworthiness}
	Table~\ref{tab:table3} links each review to its object-specific support trajectory and APCP trustworthiness degree.
	\begin{table}[H]
		\begin{center}
			\caption{Object-specific review records and trustworthiness}
			\label{tab:table3}
			\begin{adjustbox}{max width=\columnwidth}
			\begin{tabular}{|c|c|p{3.2cm}|c|c|}
				\hline
				Object & Trajectory & Review comment & Stars & Trust $\nu_{4,2}$\\
				\hline
				$M$ & $(3,0)$ & Bad product & 1 & $0$\\
				$N$ & $(2,1)$ & Not so good product & 2 & $2/3$\\
				$V$ & $(1,2)$ & Relatively good product & 3 & $2/3$\\
				$Z$ & $(0,3)$ & Premium product & 5 & $0$\\
				\hline
			\end{tabular}
			\end{adjustbox}
		\end{center}
	\end{table}
	\begin{itemize}
		\item[(i)] 	\textbf{For M:} $\boldsymbol\omega_M=(3,0)$ has modal difference $3$, so $\nu_{4,2}(\boldsymbol\omega_M)=0$. The path moves from maximum revealed support to zero support.
		\item[(ii)] 	\textbf{For N:} $\boldsymbol\omega_N=(2,1)$ has modal difference $1$, so $\nu_{4,2}(\boldsymbol\omega_N)=2/3\approx0.67$. Support decreases but remains positive in both periods.
		\item[(iii)] 	\textbf{For V:} $\boldsymbol\omega_V=(1,2)$ also has modal difference $1$, so $\nu_{4,2}(\boldsymbol\omega_V)=2/3\approx0.67$. Support increases and remains positive in both periods.
		\item[(iv)] 	\textbf{For Z:} $\boldsymbol\omega_Z=(0,3)$ has modal difference $3$, so $\nu_{4,2}(\boldsymbol\omega_Z)=0$. The path moves from zero support to maximum revealed support.
	\end{itemize}
	Thus $\{M,Z\}$ and $\{N,V\}$ form two APCP credibility classes, with $\{M,Z\}\prec_{\mathrm{cred}}\{N,V\}$. The equality within each class follows from the symmetry of modal difference; it should not be read as equality of trajectory direction.

	\subsection{Trust-Adjusted Review Scores}
	To separate favorability from behavioral credibility, let $s_x\in\{1,\ldots,5\}$ be the star rating for object $x$ and define
	\begin{equation}\label{eq:trust-adjusted-score}
		q_x^{\mathrm{adj}}=\frac{s_x}{5}\nu_{4,2}(\boldsymbol\omega_x).
	\end{equation}
	For the four records in Table~\ref{tab:table3},
	\begin{equation}
		\left[q_M^{\mathrm{adj}},q_N^{\mathrm{adj}},q_V^{\mathrm{adj}},q_Z^{\mathrm{adj}}\right]
		=\left[0,\frac{4}{15},\frac{2}{5},0\right].
	\end{equation}
	Although $Z$ has the highest raw star rating, its adjusted score is zero under the proposed rule. Among the four example records, $V$ has the largest trust-adjusted score.
	
	\section{Discussion}
	\subsection{Measure of trustworthiness and Revenue Management}
	Online review platforms affect perceived product value and, therefore, pricing, demand, and platform revenue. The proposed trustworthiness measure complements existing rating aggregates by discounting reviews associated with large changes in the reviewer’s revealed-support trajectory. In settings where platforms can observe comparison, browsing, wish-list, cart, or purchase histories, the rationality pattern can be computed from system records. Equation~\eqref{eq:trust-adjusted-score} then provides one transparent way to construct a trust-adjusted rating. A zero trust score nullifies the adjusted rating regardless of the number of stars, whereas reviews with positive trust retain a proportional contribution.

	\subsection{Extension Beyond Four Objects and Two Periods}
	The construction is not restricted to the numerical example. Let $\mathcal X$ be any finite universe with $n=\lvert\mathcal X\rvert$, and let $t=1,\ldots,T$ index the periods. An injective rank function
	\begin{equation}
		r_t:\mathcal X\longrightarrow\{0,1,\ldots,n-1\}
	\end{equation}
	uses larger values for more-preferred objects and induces $x\succ_t y$ whenever $r_t(x)>r_t(y)$. The resulting revealed-support count and trajectory are
		\begin{equation}
			d_t(x)=\left|\{y\in\mathcal X:x\succ_t y\}\right|,
			\qquad
			\boldsymbol\omega_x=(d_1(x),\ldots,d_T(x)).
		\end{equation}
	Equation~\eqref{eq:apcp-general} then applies directly to arbitrary finite $n$ and $T$. The seven-bar multiplicity profile $[1,2,3,4,3,2,1]$ is specific to $n=4$ and $T=2$, but it is only a combinatorial illustration; it is not required to recalibrate the generalized APCP score. When availability varies across periods, the missing-state symbol $\bot$ should be stored separately, and APCP should be computed only after specifying how missing observations are handled.
	
	\subsection{Conclusions and Limitations}
	The paper develops an object-specific rationality-pattern function and an APCP degree that links a review to the reviewer’s revealed sequential choices. In the worked example, the exact normalized degrees are $0$ for $M$, $Z$, and $2/3$ for $N$ and $V$. The method also distinguishes object-specific credibility from the reviewer’s aggregate inter-period consistency: the example rankings have complete pairwise reversal, even though two individual objects receive positive APCP degrees.

	The current framework has four main limitations:
	\begin{enumerate}[label=(\roman*)]
		\item it models a single selected object rather than a bundle or set-valued final choice;
		\item it follows one decision-maker and does not model household, shared-account, or multi-agent purchases;
		\item the generalized APCP formula requires empirical validation before it is used as a calibrated measure of review credibility;
		\item varying product availability introduces missing states $\bot$, which require an explicit missing-data rule before trajectories from different periods can be compared.
	\end{enumerate}
	The finite-universe and multi-period definitions remove the structural four-object restriction, but empirical calibration and validation of the generalized trust score remain necessary.
	\section*{Disclosure of Interest}
	\noindent \textbf{Declarations}\\
	\textbf{Ethical approval and consent to participate}\\
	This article does not contain studies with human participants or animals performed by any authors. Hence, ``Not applicable.''\\
	\textbf{Consent for publication}: Not applicable.\\
	Availability of data and material: No data were generated or analyzed in the manuscript. Hence, ``Not applicable.''\\
	\textbf{Competing interests}: I certify that I have no affiliations with or involvement in any organization or entity with any financial interest (such as honorarium, educational grants, participation in speaker bureaus, APCP, employment, consultancies, stock ownership, or other equity interest; and expert testimony or patent-licensing arrangements), or non-financial interest (such as personal or professional relationships, affiliations, knowledge, or beliefs) in the subject matter or materials discussed in this manuscript.\\
	\textbf{Funding}: This study was not funded by any organization.\\
	Authors' contributions: This is a single-authored article written solely by the author.\\
	\textbf{Acknowledgments}: This article was written with the help of research done by other scholars, as cited in the article. \\
	\section*{Declaration of generative AI and AI-assisted technologies in the manuscript preparation  process}
	The idea of the present manuscript is original and written by the author himself. Nothing has been generated by AI. During the preparation of this work, the author(s) used generative AI and AI-assisted tools for English-language editing and proofreading, and for assistance in generating code snippets. After using these tools, the author(s) reviewed and edited the content as needed and assume(s) full responsibility for the content of the published article.
	\section*{Software Availability and Reproducibility}
	To support the findings presented in this paper, we have developed the \textit{Review Trustworthiness Engine}, a standalone HTML-based tool that implements the models and algorithms discussed in the paper. The tool can be accessed directly in a web browser at:
	\begin{itemize}
		\item Tool: \url{https://dipankar3das.github.io/Review_Trustworthiness_Engine.html/Review_Trustworthiness_Engine.html}
	\end{itemize}
	The full source code and accompanying documentation are available in the following public repository.
	\begin{itemize}
		\item Code repository (for editors/reviewers): \url{https://github.com/dipankar3das/Review_Trustworthiness_Engine.html}
	\end{itemize}
	This resource is provided to facilitate transparency, reproducibility, and interactive exploration of the review trustworthiness framework proposed in the manuscript.

	\begingroup
	\setstretch{1.0}
	\setlength{\bibsep}{0pt}  
	\bibliographystyle{IEEEtranN}
	\bibliography{myreferences}
	\endgroup

\end{document}